\documentclass[lettersize,journal]{IEEEtran}
\usepackage[T1]{fontenc}
\IEEEoverridecommandlockouts
\usepackage{CJK}
\usepackage{subfigure}
\usepackage{colortbl}
\usepackage{bm}
\usepackage{graphicx}  
\usepackage{url}       
\usepackage{comment}
\usepackage{amsmath}
\usepackage{cite}
\usepackage{upgreek}

\usepackage{amsfonts,amssymb}

\usepackage{stfloats}
\usepackage{epstopdf} 

\usepackage{cases}
\usepackage{algorithm}
\usepackage{multirow}
\usepackage{algorithmic}
\usepackage{float}
\newcommand{\ls}[1]
{\dimen0=\fontdimen6\the\font
	\lineskip=#1\dimen0
	\advance\lineskip.5\fontdimen5\the\font
	\advance\lineskip-\dimen0
	\lineskiplimit=.9\lineskip
	\baselineskip=\lineskip
	\advance\baselineskip\dimen0
	\normallineskip\lineskip
	\normallineskiplimit\lineskiplimit
	\normalbaselineskip\baselineskip
	\ignorespaces
}

\graphicspath{{Performance_Matlab_1018/}{Simulation-1213/}{Converage/}}

\begin{document}

\title{Multiple Access Integrated Adaptive Finite Blocklength for Ultra-Low Delay in\\ 6G Wireless Networks}

\author{
	\IEEEauthorblockN{Yixin Zhang, \emph{Student Member}, \emph{IEEE}, Wenchi Cheng, \emph{Senior Member}, \emph{IEEE}, and Wei Zhang, \emph{Fellow}, \emph{IEEE}}
	\thanks{Manuscript received January 9, 2022; revised April 10, 2022.}

	\thanks{
		Part of this work was presented in IEEE Global Communications Conference, 2022~\cite{Globecom22}. 
		This work was supported in part by National Key R\&D Program of China under Grants 2021YFC3002102 and 2020TFA0711400. Manuscript received April 19, 2021; revised August 16, 2021.
		
		Yixin Zhang and Wenchi Cheng are with State Key Laboratory of Integrated Services Networks, Xidian University, Xi'an, 710071, China (e-mails: yixinzhang@stu.xidian.edu.cn; wccheng@xidian.edu.cn). 
		
		Wei Zhang is with School of Electrical Engineering and Telecommunications, the University of New South Wales, Sydney, Australia (e-mails: w.zhang@unsw.edu.au).
	}
}

\maketitle
\thispagestyle{empty}

\vspace{-50pt}

\begin{abstract}
Facing the dramatic increase of real-time applications and time-sensitive services, large-scale ultra-low delay requirements are put forward for the sixth generation (6G) wireless networks. To support massive ultra-reliable and low-latency communications (mURLLC), in this paper we propose an adaptive finite blocklength framework to reduce the over-the-air delay for short packet transmissions with multiple-access and delay-bounded demands. In particular, we first give the specified over-the-air delay model. Then, we reveal the tradeoff relationship among queuing delay, transmission delay, and the number of retransmissions along with the change of finite blocklength, as well as formulate the adaptive blocklength framework. Based on the adaptive blocklength framework and associated with grant-free (GF) access protocol, we formulate the average over-the-air delay minimization problem, where the blocklength can be adaptively changed in terms of transmission time interval (TTI) design and bandwidth allocation to achieve the optimal tradeoff and obtain its minimum over-the-air delay. We develop the cooperative multi-agent deep Q-network (M-DQN) scheme with a grouping mechanism to efficiently solve the average over-the-air delay minimization problem. Numerical results validate our proposed adaptive blocklength scheme outperforms corresponding schemes in long-term evolution (LTE) and the fifth generation (5G) new radio (NR).
\end{abstract}

\begin{IEEEkeywords}
Massive ultra-reliable and low-latency communications (mURLLC), delay tradeoff, adaptive finite blocklength, grant-free access, over-the-air delay minimization.
\end{IEEEkeywords}

\section{Introduction}

\IEEEPARstart{W}{ith} the rapid development of various vertical businesses in the Internet of Things (IoT), such as autonomous vehicles, factory automation, remote control, healthcare virtual, and augmented reality, stringent delay and access requirements become more and more important in the beyond fifth generation (B5G) and the sixth generation (6G) wireless networks~\cite{She1Intro,EC2}. In the fifth generation (5G) communication networks, the third generation partnership project (3GPP) presents ultra-reliable low-latency communications (URLLC) and massive machine type communications (mMTC) as the core services to provide mission-critical support and massive connectivity~\cite{Editor6}. Specifically, URLLC aims to provide end-to-end (E2E) delay of less than 1 ms for 32-bit packet transmission~\cite{38.913}. In the case of mMTC, the main challenge is to support a large number of devices with limited radio resources. 6G wireless networks serve all walks of life, which will be further expanded to support the fully intelligent connectivity networks~\cite{Intro-6G}, thus putting forward more stringent requirements on efficient, delay-bounded, and reliable communications~\cite{Intro-6G2}. Since the separate URLLC and mMTC in 5G cannot provide massive low-latency communications for various delay-sensitive services, integrating URLLC with mMTC, massive URLLC (mURLLC) has been proposed in 6G wireless networks to support large-scale real-time services and applications within the delay bound and error limit~\cite{mURLLC2}. However, in the existing network architecture, the over-the-air delay constraints of mURLLC are difficult to be satisfied. As a result, delay timeout under massive access conditions remains a difficult problem and time-saving solutions are still very important for future IoT networks.

To reduce the E2E delay and meet stringent delay demands in mURLLC, we need to comprehensively analyze each delay component contained in the E2E delay. Specifically, the E2E delay contains the over-the-air delay, the backhaul delay, and the routing delay. The backhaul delay, which depends on the physical distance and medium, is much shorter than $1$~ms with fiber backhaul~\cite{She1Intro}. As for the routing delay, software-defined routers (SDRs) have emerged as a potential solution to support relatively low routing delay with effective routing path management~\cite{Routing}. The over-the-air delay, which is one of the most important parts as well as the bottleneck of the E2E delay, contains the queuing delay, the transmission delay, the processing delay, and the propagation delay. The over-the-air delay, depending on the one time access delay and the number of retransmissions, is related to the specific access protocol~\cite{Access_URLLC}. On the one hand, different access protocols lead to different delay compositions, that is, the value of queuing delay, as well as the number of processing delays and propagation delays contained in one time access delay are different~\cite{Editor4}. On the other hand, the number of retransmissions, which impacts the over-the-air delay, is also determined by the access protocol.

Most existing works focus on separately optimizing a single component of the over-the-air delay. For queuing delay, the queuing theory was widely used for reducing the average queuing delay\cite{Queuing_Ave_1}, while the effective capacity was proposed to reduce the upper bound of queuing delay\cite{Queuing_EC_1,Queuing_EC_2}. For transmission delay, changing the subcarrier spacing (SCS), 3GPP introduced the new transmission time interval (TTI) as $0.5$~ms, $0.25$~ms, $0.125$~ms, and $0.0625$~ms in 5G new radio (NR) to reduce the transmission delay~\cite{38.211}. Considering the characteristic of IoT packets and the shortened TTI frame structure proposed in 5G NR, the long blocklength, which is used for high-capacity-demanded service, is not applicable for the delay-sensitive application with the small-size packet. As a result, finite blocklength coding (FBC) has been proposed as a promising technology to adapt to short packet transmissions with shortened TTIs and achieve extremely low transmission delay~\cite{FB,FB-MIMO}. 

To further reduce the over-the-air delay, the impact of access protocol on delay is highly required to be analyzed. In long-term evolution (LTE), grant-based (GB) access has the four-step handshake procedure consisting of scheduling request, uplink grant, uplink data transmission, as well as acknowledgment (ACK) and negative acknowledgment (NACK) transmission, resulting in a total E2E delay of $17$~ms~\cite{36.881}. GB access requires a large amount of time and signaling for the connection setup before the actual data transmission, which is very hard to efficiently support IoT traffic~\cite{Editor2}. To meet the massive low-delay requirements in mURLLC, grant-free (GF) random access was proposed in 5G NR to accommodate massive access within the delay limit~\cite{GF-WHOLE}. The grant request step has been removed in a two-step GF access procedure, where additional over-the-air delay and overhead can be avoided, thus supporting massive connectivity in mMTC with low latency and high spectrum efficiency. 

Even though various studies on the advantages and frameworks of using finite blocklength and GF access for low-delay communications have been carried out, the frame structures in LTE and 5G NR are still based on the fixed TTI~\cite{38.211}, which corresponds to the fixed blocklength. Also, only the impact of finite blocklength on transmission delay has been taken into account, without considering other parts of E2E delay~\cite{Add1,Add2}. However, in the finite blocklength regime, the transmission delay, the queuing delay, and the number of retransmissions contained in the over-the-air delay change in different directions as the finite blocklength changes. Only reducing the blocklength in 5G NR leads to the increase in the queuing delay and the number of retransmissions~\cite{Xiao,38.211}. As a result, a fixed blocklength leads to an imbalance among the above three components, resulting in an increase in the over-the-air delay. In addition, GF access leads to packet collisions, resulting in retransmissions and an increase in over-the-air delay. To fully consider the overall impact of GF access and finite blocklength on the delay performance, the corresponding GF access and finite blocklength combined analysis, as well as its corresponding adaptive blocklength framework with joint components optimization, are highly needed.

To solve the above-mentioned problem, in this paper, we propose an adaptive blocklength framework for short packet transmissions to reduce the over-the-air delay by combining finite blocklength and GF access analysis in the mURLLC system.\footnote{In this paper, we comprehensively provide the adaptive blocklength framework with joint TTI and bandwidth design, which is more challenging than the previous work~\cite{Globecom22}.} In particular, we reveal the tradeoff relationship among queuing delay, transmission delay, and the number of retransmissions along with the change of finite blocklength, based on which we formulate the adaptive blocklength framework. Then, we drive the analytical expressions of the successful access probability and the steady-state probability distribution, thus establishing the queuing state update model of the adaptive blocklength framework. On this basis, we formulate the average over-the-air delay minimization problem, where the blocklength can be adaptively changed in terms of TTI design and bandwidth allocation. We also develop a cooperative multi-agent deep Q-network (M-DQN) scheme with a grouping mechanism to efficiently deal with this non-convex problem. The optimal tradeoff is achieved and the minimum over-the-air delay is obtained. 

The rest of this paper is organized as follows. Section~\ref{sec:sys} introduces the system model. Section~\ref{sec:AFB} reveals the new delay tradeoff and proposes the adaptive blocklength framework. Section~\ref{sec:min} presents the GF-based average over-the-air delay minimization problem and develops the cooperative M-DQN based scheme. Section~\ref{sec:result} provides the numerical results. Finally, we conclude this paper in Section~\ref{sec:con}.

\section{System Model}\label{sec:sys}
In this section, we introduce the over-the-air delay model and analyze the delay components. Then, a queuing state update model based on discrete Markov chain is established.

\begin{figure}[htbp]
\centering\includegraphics[width=3.5in]{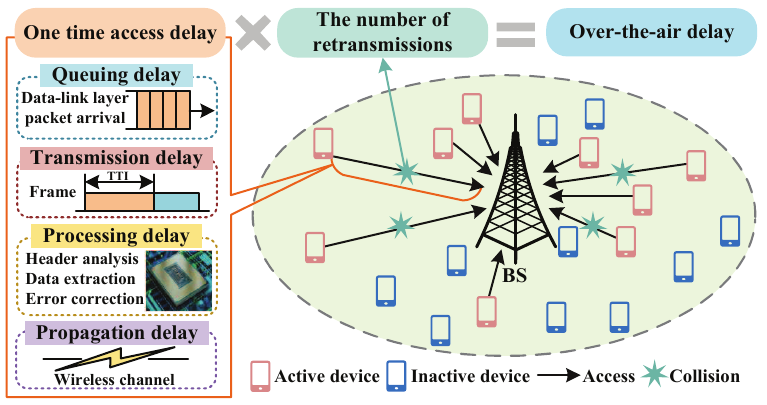}
\caption{The over-the-air delay model.}\label{fig:System}
\end{figure}

\subsection{The Over-the-Air Delay Model}
Figure~\ref{fig:System} shows the over-the-air delay model. The over-the-air delay depends on the one time access delay and the number of retransmissions. Specifically, the number of retransmissions relates to the access protocol and real-time traffic load. The one time access delay is composed of queuing delay, transmission delay, processing delay, and propagation delay. The queuing delay, which depends on the number of packets that arrived earlier and are queuing for transmission, refers to the time duration between the new packet arrival from the data-link layer and the first attempt of access. The transmission delay, which highly depends on the TTI of the frame, refers to the time duration of sending the first bit to the last bit of the packet. The processing delay, usually on the order of microseconds or less with the high-speed processor, refers to the time spent by the central processing unit handling the data, including header analysis, data extraction, and error correction. The propagation delay, which relates to the propagation distance and the propagation speed, refers to the consumed time that one bit is transmitted from the sender to the receiver on the wireless channel.

\subsection{The Transmission Model}
We consider an mURLLC system with one base station (BS) and $K$ devices, where each device experiences both large-scale fading and small-scale fading. The small-scale Rayleigh fading coefficient, denoted by $h$, is with zero mean and unit variance, i.e., $\mathbb E(|h|^2)=1$. In many applications of mURLLC, devices are either static or have low mobility \cite{Channel1}, where the duration of each frame is less than the channel coherence time. Thus, the channel is quasi-static and $h$ can be regarded as a constant over several frames. We assume that all devices use full path-loss inversion power control with a threshold $P_0$~\cite{P0}. That is, each device controls its transmit power to guarantee the signal power received at the BS is equal to a predetermined value $P_0$. Considering that it is difficult for devices to obtain instantaneous channel state information (CSI), devices use statistical CSI for inversion power control, where statistical CSI is constant over a long time period and is easy to obtain. In this way, the transmit power of the $k$-th device is $ P_k = \frac{d_k^{\alpha}}{\rho_0 |h|^2}P_0$, where $d_k$, $\alpha$, and $\rho_0$ denote the distance between the $k$-th device and the BS, the pass-loss exponent, and the channel power gain at a reference distance of $1$ meter, respectively. Thus, the signal-to-noise ratio (SNR) at the BS can be written as $\chi = \frac{P_0}{\sigma^2}$, where ${\sigma^2}$ denotes the noise power.

\subsection{The Pakcet Arrival Model}
We consider a general mURLLC scenario, where different types of devices sporadically generate short packets according to their packet arrival rates. Generally, we use the Poisson process as a traffic model~\cite{Possion}, where each device generates new packets following the Poisson distribution with its packet arrival rate $\lambda_k \ (1 \leq k \leq K)$. Thus, the probability mass function (PMF) of newly arrived packets generated by the $k$-th device, denoted by $p^{\rm gen}_{k}$, can be expressed as
\begin{equation}
	p^{\rm gen}_{k}\left(a_k\right)= \exp\left({-\lambda_k T_{\max}}\right) \frac{\left(\lambda_k T_{\max} \right)^{a_k}}{a_k!},
\end{equation}
where $T_{\max}$ represents the period duration and $a_k$ represents the number of packets generated by the $k$-th device within the given time interval.

\subsection{The Queuing State Update Model}
According to the packet arrival rate $\lambda_k \ (1 \leq k \leq K)$, the maximum queuing length of the $k$-th device can be derived based on the number of arrival packets as follows:
\begin{equation}
	M^{\rm Que}_k = \left\lceil \sum_{i=0}^{\infty}i  p^{\rm gen}_{k}(i) \right\rceil, \label{maximum_queuing_length}
\end{equation}
where the operation $\lceil \cdot \rceil$ presents the round-up operation. Then, the behavior of each device can be modeled as a discrete Markov chain process~\cite{Markov}. Each state represents the queuing length of each device, where the queuing length implies the number of packets waiting in the queue. The state space of the queuing state update model, denoted by $\mathcal{M}$, can be expressed as 
\begin{equation}
	\mathcal{M}=\left\{0,\cdots, m ,\cdots,M^{\rm Que}_k\right\}.
\end{equation}
When the number of packets exceeds the threshold $M^{\rm Que}_k$, excess packets are discarded to meet the low-delay requirements. At the beginning of the period duration, the initial state depends on the initial packet arrival rate $\lambda_k \ (1 \leq k \leq K)$. A state transition happens whenever each device attempts to perform the transmission, which results in two cases, one is success while the other is failure with unexpected collisions or errors. In the queuing state update model, state $0$ indicates the idle state that the device has no packets to transmit, while state $m \ (1\leq m \leq M^{\rm Que}_k)$ indicates the active state that the queue is with $m$ packets waiting for transmission. The state probability of the $k$-th device at time $t$, denoted by $\boldsymbol\pi_k(t)$, can be expressed as 
\begin{equation}
	\boldsymbol\pi_k(t)=\left\{\pi_{k,0}(t),\cdots, \pi_{k,m}(t), \cdots, \pi_{k,M^{\rm Que}_k}(t)\right\},
\end{equation}
where $\pi_{k,m}(t) \ (0\leq m \leq M^{\rm Que}_k)$ represents the state probability that the queuing length of the $k$-th device is equal to $m$ at time $t$.

\section{Adaptive Blocklength Framework}\label{sec:AFB}
For short packet transmission in the mURLLC scenario, the infinite blocklength based on Shannon capacity is no longer applicable. As a result, we analyze the finite blocklength and propose an adaptive blocklength framework to further increase the design flexibility.
\subsection{The Blocklength Structure}
In the finite blocklength regime, the blocklength, which refers to the number of symbols transmitted in a block, denoted by $n$, can be expressed as
\begin{align}\label{eq:TB}
	n=TW,
\end{align}
where $T$ represents the time span, i.e., the TTI, and $W$ represents the bandwidth resource occupied by the current block. Using the normal approximation \cite{FB2}, the achievable rate in the finite blocklength regime, denoted by $R(n)$, can be approximated as 
\begin{align}\label{eq:FBR}
	R(n) \approx {\log _2}\left(1 + \chi \right) - \sqrt {\frac{V}{n}} Q^{ - 1}(\varepsilon){\log _2}e,
\end{align}
where $V=1-\left(1+\chi\right)^{-2}$ denotes the channel dispersion, $\varepsilon$ denotes the packet error probability, and $Q^{-1}(\varepsilon)$ denotes the inverse of Q function with $ Q(\varepsilon) = \int_{\varepsilon}^{\infty} \frac{1}{\sqrt{2\pi}}\exp \left( -\frac{u^2}{2} \right)\mathrm{d}u$. For the finite blocklength, a significant packet error probability is introduced. According to \cite{FB}, the packet error probability is given by
\begin{equation}\label{eq:Error1}
	\varepsilon=Q\left(\sqrt{\frac{n}{V}}\frac{\log_2(1+\chi)-R(n)}{\log_2e}\right).
\end{equation}
 
\subsection{The Tradeoff Relationship}
\begin{figure}[htbp]
	\centering
	\includegraphics[scale = 0.8]{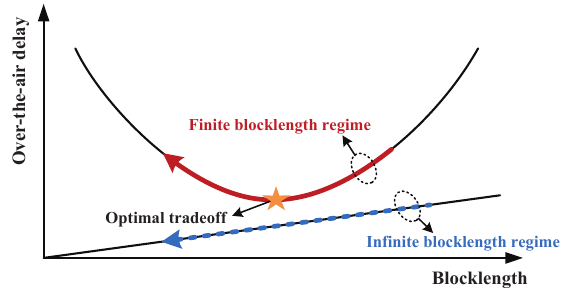}
	\caption{The over-the-air delay versus the blocklength in the case of finite blocklength and infinite blocklength.}
	\label{Tradeoff}
\end{figure}
Figure~\ref{Tradeoff} shows the over-the-air delay versus the blocklength in the case of finite blocklength and infinite blocklength. In the infinite blocklength regime, the achievable rate, as well as the Shannon capacity, does not change with the blocklength. Hence, the service rate of queuing process in the link layer keeps unchanged, where the corresponding queuing delay is also unchanged. At the same time, the change of blocklength does not affect the number of retransmissions and only the transmission delay decreases as the blocklength decreases. As a result, the over-the-air delay monotonously decreases as the blocklength decreases, which means shortening the blocklength can reduce the over-the-air delay. 

In the finite blocklength regime, the transmission delay gradually decreases as the blocklength decreases. However, according to~(\ref{eq:FBR}), as the blocklength decreases, the achievable rate decreases, correspondingly resulting in an increase of queuing delay. In addition, based on~(\ref{eq:Error1}), the packet error probability, as well as the number of retransmissions, increase as the blocklength decreases. As the blocklength decreases, the over-the-air delay no longer monotonically decreases, which means there is a tradeoff relationship among the transmission delay, the queuing delay, and the number of retransmissions. When the transmission delay, the queuing delay, and the number of retransmissions contained in the over-the-air delay reach the optimal tradeoff, the minimum over-the-air delay is achieved. Therefore, in the following we propose an adaptive blocklength framework to flexibly change the blocklength of each packet based on network traffic load to reach the optimal tradeoff point and minimize the over-the-air delay. 

\subsection{The Adaptive Blocklength Framework}
\begin{figure}[htbp]
	\centering
	\includegraphics[scale = 0.8]{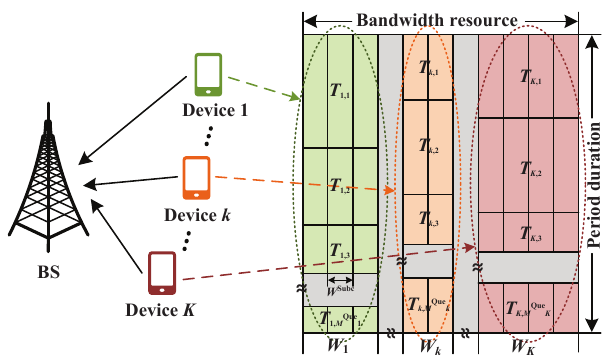}
	\caption{Adaptive blocklength framework with different TTIs and bandwidths.}
	\label{Multiuser}
\end{figure}
Figure~\ref{Multiuser} shows the adaptive blocklength framework with different TTIs and bandwidths. Since the blocklength is obtained by the TTI and bandwidth based on~(\ref{eq:TB}), we need to jointly change the two factors to generate the corresponding adaptive blocklength. Specifically, different from the fixed TTI structure in LTE and 5G NR, we set a variable TTI for each packet of each device. In addition, based on the real-time traffic load, we allocate the bandwidth, i.e., assign subchannels to different devices. We assume that the number of subchannels is set to $M^{\rm Subc}$, where the orthogonal bandwidth allocation scheme is employed. The bandwidth allocation indicator, which indicates whether the $c$-th subchannel is allocated to the $k$-th device, denoted by $\mathcal {J}(c,k)$, can be written as
\begin{equation}
	\mathcal {J}(c,k) = \left\{
	\begin{aligned}
		&1, & & {\mbox {if}} \ \  c \in \mathcal{W}_k, \\
		& 0, & & {\mbox {otherwise}},
	\end{aligned}
	\right.
\end{equation}
where $\mathcal{W}_k$ denotes the set of subchannels allocated to the $k$-th device. In addition, the bandwidth allocation indicator $\mathcal{J}(c, k)$ needs to satisfy
\begin{align}\label{eq:indicator_constrain}
	\sum\limits_{k = 1}^K \mathcal{J}(c, k) = 1, \ 1 \leq c \leq M^{\rm Subc}.
\end{align} 
Therefore, the bandwidth of the $k$-th device, denoted by $W_k$, can be expressed as
\begin{equation}
	W_k = \sum_{c=1}^{M^{\rm Subc}}\mathcal {J}(c,k)  W^{\rm Subc},
\end{equation}
where $W^{\rm Subc}$ denotes the bandwidth per subchannel. As for TTI, it can be continuously changed in each period according to the actual traffic load. Thus, the blocklength of the $m$-th packet from the $k$-th device, denoted by $n_{k,m}$, can be expressed as
\begin{equation}
	n_{k,m} = T_{k,m}  W_k ,
\end{equation}
where $T_{k,m}$ denotes the TTI of the $m$-th packet from the $k$-th device. Specifically, the adaptive blocklength vector, denoted by $\boldsymbol n $, can be expressed as
\begin{equation}
	\begin{split}
		\boldsymbol n = 
		&\Big\{
		\underbrace{T_{1,1} W_1, \cdots, T_{1,{M^{\rm Que}_1}} W_1}_{\boldsymbol n_1}, \cdots , 
		\underbrace{T_{k,1} W_k, \cdots, T_{k,{M^{\rm Que}_k}} W_k}_{\boldsymbol n_k},\\
		&\cdots , \underbrace{T_{K,1} W_K, \cdots, T_{K,{M^{\rm Que}_K}} W_K}_{\boldsymbol n_K}
		\Big\},
	\end{split}
\end{equation}
where ${\boldsymbol n_k}$ represents the adaptive blocklength vector of the $k$-th device. By jointly considering the TTI and bandwidth, the adaptive blocklength framework with different TTIs and bandwidths thus forms and is shown in Fig.~\ref{Multiuser}. The total blocklength of the $k$-th device, denoted by $n_k$, can be expressed as
\begin{equation}
	n_k = \sum_{m=0}^{M^{\rm Que}_k} \sum_{c=1}^{M^{\rm Subc}} \mathcal {J}(c,k)  T_{k,m}  W^{\rm Subc}.
\end{equation}


\subsection{The Analysis of Adaptive Blocklength Framework}
To meet the massive low-delay requirements in mURLLC, GF access is adopted in the adaptive blocklength framework. Taking GF access into account, we analyze the corresponding parameters in the adaptive blocklength framework, including packet collision probability, packet error probability, successful access probability, and steady-state probability distribution of the queuing state update model. 

\subsubsection{The Packet Collision Probability}
In GF access, once a device becomes active, it directly transmits data with a preamble. We assume the active devices first randomly select a preamble from the preamble pool containing $M^{\rm Pre}$ preambles. In the mURLLC scenario, since a large number of devices sporadically and randomly send short packets only during the active phase, reserving preambles for each device in advance results in a large preamble overhead~\cite{Preamble}. As a result, we assume the number of preambles is finite with $M^{\rm Pre} < K$, which can lead to packet collisions in the preamble selection stage. The packet collision occurs if more than one active device selects the same preamble. 
The number of devices selecting the $s$-th preamble in the preamble pool, denoted by $M^{\rm Pre}_s$, approximately follows Binomial distribution with parameters $\left (K,\frac{1}{M^{\rm Pre}}\right)$. It holds that the variables $M^{\rm Pre}_s\ (s = 1, 2,\cdots, M^{\rm Pre})$ are mutually independent. The device successfully accesses by selecting the $s$-th preamble if only the $s$-th preamble is selected only by this device. We denote the probability of the event that the $s$-th preamble is selected only by one device as $p^{\rm one}_s$, which can be calculated as
\begin{equation}
	p^{\rm one}_s = \Pr\left\{M^{\rm Pre}_s = 1\right\} = \left(\frac{1}{M^{\rm Pre}}\right) \left(1-\frac{1}{M^{\rm Pre}}\right)^{K-1}.
\end{equation}
Therefore, the probability of a device selecting any preamble that is not selected by other devices, denoted by $p^{\rm one}$, can be calculated as
\begin{equation}
	p^{\rm one} = \sum_{s = 1}^{M^{\rm Pre}} p^{\rm one}_s =  \left(1-\frac{1}{M^{\rm Pre}}\right)^{K-1}.
\end{equation}

\subsubsection{The Packet Error Probability}
Since the packet error probability exists in the finite blocklength, it severely impacts the over-the-air delay performance. In other words, even if a device becomes active and the packet collision is avoided within the block, the transmitted packet may still not be successfully received by the BS. Because it is difficult to directly deal with Q function $Q(\varepsilon)$ in (\ref{eq:Error1}), we use a linear function to tightly approximate Q function according to~\cite{pe_appro,pe_appro_new}. Thus, the packet error probability of the $m$-th packet from the $k$-th device can be rewritten as
\begin{equation}
	\varepsilon_{k,m} \approx
	\begin{cases}
		1,& \chi \leq \tau_{k,m}^1,\\
		\frac{1}{2}-\mu_{k,m}  \left (\chi-\xi_{k,m}\right),& \tau_{k,m}^1 < \chi \leq \tau_{k,m}^2,\\
		0,& \chi > \tau_{k,m}^2,\\
	\end{cases}
\end{equation}
where $\mu_{k,m} = \frac{1}{2 \pi }\sqrt{\frac{n_{k,m}}{\left(2^{2B/n_{k,m}}-1\right)}}$, $\xi_{k,m} = 2^{B/n_{k,m}} - 1$, $\tau_{k,m}^1 = \xi_{k,m} - \frac{1}{2\mu_{k,m}}$, $\tau_{k,m}^2 = \xi_{k,m} + \frac{1}{2\mu_{k,m}}$, and $B$ denotes the number of bits contained in one short packet. Therefore, the packet error probability of the $m$-th packet from the $k$-th device, denoted by $p^{\rm err}_{k,m}$, can be rewritten as follows:
\begin{equation}\label{Eq:pac-err}
	\begin{split}
		p^{\rm err}_{k,m} = &\mathbb{E}[\varepsilon_{k,m}]\\
		= & \int_{0}^{\tau_{k,m}^1} f_{\chi}(x) \mathrm{d}x \\ &\qquad+\int_{\tau_{k,m}^1}^{\tau_{k,m}^2} \left( \frac{1}{2}-\mu_{k,m} (x-\xi_{k,m}) \right) f_{\chi}(x) \mathrm{d}x \\
		= & 1-\mu_{k,m}\frac{P_0}{\sigma^2}\Bigg(\exp{\bigg[-\frac{\sigma^2}{P_0}\left(\xi_{k,m} - \frac{1}{2\mu_{k,m}}\right)\bigg]}\\		
		&\qquad\qquad -\exp{\bigg[-\frac{\sigma^2}{P_0}\left(\xi_{k,m} + \frac{1}{2\mu_{k,m}}\right)}
		\bigg]\Bigg),
	\end{split}
\end{equation}	
where $f_{\chi}(x) =  \frac{\sigma^2}{P_0} \exp {\left(-\frac{\sigma^2 x}{P_0}\right)}$ for $x \geq 0$ is the probability distribution function (PDF) of the SNR.

\begin{figure}[hbp]
	\centering\includegraphics[width=3.5in]{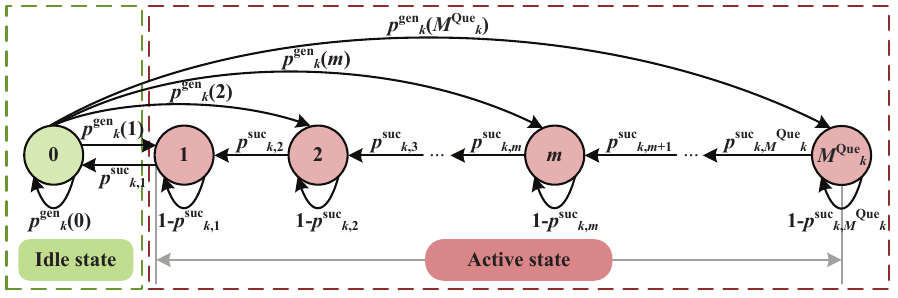}
	\caption{The queuing state update model of GF access based adaptive blocklength framework.}\label{fig:MarkovChain}
\end{figure}

\subsubsection{The Successful Access Probability}
The packet can successfully access the BS only when it selects the unique preamble without the transmission error~\cite{Similar3}. Therefore, the successful access probability of the $m$-th packet from the $k$-th device, denoted by $p^{\rm suc}_{k,m}$, can be derived as
\begin{equation}
	p^{\rm suc}_{k,m} = p^{\rm one} \left(1- p^{\rm err}_{k,m}\right).
\end{equation}

\subsubsection{The Steady-State Probability}
In order to mathematically derive the expression of over-the-air delay, we should obtain the stationary distribution of each device. The stationary distribution of the $k$-th device, denoted by $\boldsymbol\pi_k = \{\pi_{k,0}, \cdots, \pi_{k,m}, \cdots, \pi_{k,M^{\rm Que}_k}\}$, can be written as follows: 
\begin{equation}
	\boldsymbol\pi_k=\lim\limits_{t\to \infty}\boldsymbol \pi_k(t)=\lim\limits_{t\to \infty} \boldsymbol P_k ^t \pi_k(0),
\end{equation}
where $\boldsymbol P_k$ represents the one-step state transition matrix and $\pi_k(0)$ represents the initial state probability distribution of the $k$-th device. Fig.~\ref{fig:MarkovChain} shows the queuing state update model of GF-ased adaptive blocklength framework. Based on the adaptive blocklength $\boldsymbol{n}$, each state has a unique transition probability in the queuing state update model. To obtain the steady-state probability distribution, the steady-state probability function set of the $k$-th device can be expressed as follows:
\begin{equation}\label{eq:steady_function_set}
	\left\{
	\begin{aligned}
		&\sum_{m=0}^{M^{\rm Que}_k}\pi_{k,m} = 1, & & \\
		& \pi_{k,0} \left(\sum_{m=0}^{M^{\rm Que}_k}p^{\rm gen}_k(m)\right) = \pi_{k,0}  p^{\rm gen}_k(0) + \pi_{k,1}  p^{\rm suc}_{k,1},&&\\
		&\pi_{k,m} = \pi_{k,0} p^{\rm gen}_k(m) + \pi_{k,m} \left(1-p^{\rm suc}_{k,m}\right)  \\ && \hspace{-7cm}+ \pi_{k,m+1}  p^{\rm suc}_{k,m+1},  
		1 \leq m < M^{\rm Que}_k,&&\\
		&\pi_{k,M^{\rm Que}_k} = \pi_{k,0} p^{\rm gen}_k\left(M^{\rm Que}_k\right) + \pi_{M^{\rm Que}_k} \left(1-p^{\rm suc}_{k,M^{\rm Que}_k}\right), &&
	\end{aligned}
	\right.
\end{equation} where $\pi_{k,m}$ represents the steady-state probability of the $m$-th packet from the $k$-th device in the queue. After simplifying the function set~(\ref{eq:steady_function_set}), the steady-state probability of the $k$-th device can be expressed as (21). 
\begin{figure*}[ht] 
	\centering
	\hrulefill
	\begin{equation}\label{Eq:state}
		\left\{
	\begin{aligned}
		&\pi_{k,0} = \frac{ \prod_{i=1}^{M^{\rm Que}_k} p^{\rm suc}_{k,i}}{ \prod_{i=1}^{M^{\rm Que}_k} p^{\rm suc}_{k,i} + \sum_{j=1}^{M^{\rm Que}_k} \Bigg[\left(\prod_{r=1,r \neq j}^{M^{\rm Que}_k} p^{\rm suc}_{k,r}\right) \left(\sum_{l=j}^{M^{\rm Que}_k} p^{\rm gen}_k(l) \right) \Bigg] }, \ 1 \leq k \leq K, & & \\
		& \pi_{k,m} = \frac{\pi_{k,0} \left(\sum_{l=m}^{M^{\rm Que}_k}p^{\rm gen}_k(l)\right)}{p^{\rm suc}_{k,m}}, \ 1 \leq k \leq K, \ 1 \leq m \leq M^{\rm Que}_k.&&
	\end{aligned}
	\right.
	\end{equation}
\hrulefill
\end{figure*}
It can be seen from~(\ref{Eq:state}) that the steady-state probability distribution is a function of the packet arrival rate and the successful access probability of each device at each attempt.

\section{The Over-the-Air Delay Minimization Problem}\label{sec:min}
To satisfy stringent delay demands in mURLLC scenario, we propose the average over-the-air delay minimization problem that considers the queuing delay, the transmission delay, and the number of retransmissions to reduce the over-the-air delay by adaptively changing the blocklength of each packet.
\subsection{The Delay Analysis of GF Access}
\begin{figure}[htbp]
	\centering\includegraphics[width=3in]{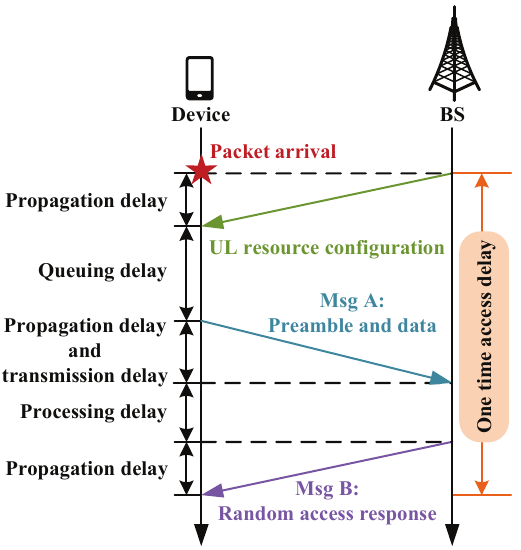}
	\caption{The over-the-air delay components of GF access.}\label{fig:GF}
\end{figure}
Figure~\ref{fig:GF} shows the over-the-air delay components of GF access. Different from GB access, the active device does not need to wait for the scheduling grant from BS in GF access. That is, once a device becomes active, it directly transmits data with a preamble and waits for the ACK from BS, which can omit the grant step to reduce delay. For contention resolution (CR) and delay limit, the active device starts a CR timer when transmitting a packet. If an ACK is received before the CR timer expires, a successful packet transmission of the device is finished.

As shown in Fig.~\ref{fig:GF}, we consider the over-the-air delay as the time interval from each request of the active device is generated until it is successfully transmitted to the BS. The one time access delay in GF access, which means the packet is successfully transmitted at once without any packet collision, denoted by $D^{\rm Acc}$, can be expressed as

\begin{equation}
	D^{\rm Acc} = D^{\rm Que} + D^{\rm Tra}  + 3D^{\rm Prop} + D^{\rm Proc},
\end{equation}
where $D^{\rm Que}$, $D^{\rm Tra}$, $D^{\rm Prop}$, and $D^{\rm Proc}$ denote the queuing delay, the transmission delay, the propagation delay, and the processing delay, respectively. GF access leads to potential collisions because the preambles are no longer reserved for a specific packet. When the packet experiences collisions, both Step 1 (Msg A) and Step 2 (Msg B) should be repeated until the packet is successfully transmitted. If collisions happen, the over-the-air delay, denoted by $D^{\rm Ota}$, can be obtained as follows:
\begin{equation}
	D^{\rm Ota} = M^{\rm Re}  D^{\rm Acc},
\end{equation}
where $M^{\rm Re}$ represents the number of retransmissions.

\subsection{The Average Over-the-Air Delay Minimization Problem Formulation}
Based on the above adaptive blocklength framework, we can reformulate the expression of the over-the-air delay. The queuing delay of the $k$-th device, denoted by $D^{\rm Que}_k$, can be expressed as
\begin{equation}
	D^{\rm Que}_k = \sum_{m=1}^{M^{\rm Que}_k}  \left(\pi_{k,m}  D^{\rm Que}_{k,m} \right),\  1 \leq k \leq K,
\end{equation} where $D^{\rm Que}_{k,m}$ represents the queuing delay of the $m$-th packet from the $k$-th device. 
It can be written as
\begin{equation}
\begin{split}	
	D^{\rm Que}_{k,m} = \sum_{l=0}^{m-1} \bigg(\Big(T_{k,l}  +  & D^{\rm P}\Big)  \mathbb{E}\left[M^{\rm Re}_{k,l}\right]\bigg),\\ & 1 \leq k \leq K, \  1 \leq m \leq M^{\rm Que}_k,
\end{split}
\end{equation} where $D^{\rm P}=3D^{\rm Prop} + D^{\rm Proc}$, $\mathbb{E}\left[ \cdot \right]$ represents the expectation, and $M^{\rm Re}_{k,l}$ represents the number of retransmissions until the $l$-th packet of the $k$-th device is successfully transmitted. $M^{\rm Re}_{k,l}$ follows a geometric distribution with the parameter $p^{\rm suc}_{k,l}$~\cite{Stochastic}. Thus, the PMF of $M^{\rm Re}_{k,l}$ can be expressed as follows:
\begin{equation}
	f_{M^{\rm Re}_{k,l}}(x)=p^{\rm suc}_{k,l} \left(1-p^{\rm suc}_{k,l}\right)^{x-1}, \  x\in\{1,2,3,\cdots\}.
\end{equation}
The transmission delay of the $k$-th device, denoted by $D^{\rm Tra}_k$, can be written as
\begin{equation}
	D^{\rm Tra}_k =  \sum_{m=0}^{M^{\rm Que}_k}\Big(T_{k,m}  \mathbb{E}\left[M^{\rm Re}_{k,m}\right]\Big), \ 1 \leq k \leq K.
\end{equation}The processing delay and the propagation delay of the $k$-th device, denoted by $D^{\rm PP}_k$, can be expressed as
\begin{equation}
	D^{\rm PP}_k =  \sum_{m=0}^{M^{\rm Que}_k}\Big( D^{\rm P}  \mathbb{E}\left[M^{\rm Re}_{k,m}\right]\Big) , \ 1 \leq k \leq K.
\end{equation}Therefore, the over-the-air delay of the $k$-th device, denoted by $D^{\rm Ota}_k$, can be expressed as
\begin{equation}
	\begin{aligned}
	D^{\rm Ota}_k &= D^{\rm Que}_k + D^{\rm Tra}_k + D^{\rm PP}_k , \ 1 \leq k \leq K,  \\
	\end{aligned}
\end{equation}
where $D^{\rm Ota}_k$ is the function of blocklength $\boldsymbol n_k$. In this paper, we aim to minimize the average over-the-air delay of $K$ devices. Thus, the average over-the-air delay minimization problem, denoted by $\textbf{\textit{P}1}$, can be expressed as
\begin{align}
	\textbf{\textit{P}1:\ } &\min_{\boldsymbol n }\ \ D^{\rm Ota}_{\rm Ave}= \frac{1}{K}\sum_{k=1}^{K} D^{\rm Ota}_k \\
	\  \mathrm{s.t.}\ \  &1).\ \sum_{m=0}^{M^{\rm Que}_k} T_{k,m} \leq T_{\max}, \ 1 \leq k \leq K,\;\label{Eq:constranit1}\\
	\ \  &2).\ \sum_{k=1}^{K}\mathcal {J}(c,k) = 1, \ 1 \leq c \leq M^{\rm Subc},\label{Eq:constranit3}\\
	\ \  &3).\ \sum_{c=1}^{M^{\rm Subc}} \mathcal {J}(c,k) \sum_{m=0}^{M^{\rm Que}_k}  \Big(R(T_{k,m}W_k)  T_{k,m}  W^{\rm Subc}  \Big) \nonumber \\& \qquad \quad\geq \sum_{a_k=0}^{M^{\rm Que}_k} \Big( p^{\rm gen}_{k}\left(a_k\right)  a_k \Big)  B, \ 1 \leq k \leq K, \label{Eq:constranit4}
\end{align}where $D^{\rm Ota}_{\rm Ave}$ represent the average over-the-air delay of $K$ devices. The constraint (\ref{Eq:constranit4}) represents that the amount of data can be transmitted is larger than the total number of bits brought by the arrived packets. Since $\textbf{\textit{P}1}$ is a non-convex problem with joint TTI-bandwidth design and continuous-discrete mixed constraints, it is hard to transform $\textbf{\textit{P}1}$ into multiple solvable convex sub-problems and find out the optimal solution. Deep reinforcement learning (DRL) is proposed as a promising method to solve the complex and non-convex problem in URLLC~\cite{DRL-URLLC-2,DRL-URLLC-3,DRL-URLLC-4}. In addition, the collaboration between massive devices should also be fully considered in the mURLLC scenario. Thus, in the following, we propose a multi-agent DRL method to solve this problem.

\subsection{Problem Transformation}
In this section, we provide a DRL based scheme for the average over-the-air delay minimization problem to design TTI and allocate bandwidth. Generally speaking, DRL can be described by a Markov Decision Process (MDP). The MDP is defined as a tuple $\left(\mathcal{S}, \mathcal{A}, r, \mathcal{P}\right)$, where $\mathcal{S}$ is the state space set, $\mathcal{A}$ is the action space set, $r$ is the instant reward, and $\mathcal{P}$ is the mapping probability of transferring from a current state to a new state after taking an action. 

Considering the characteristic of mURLLC scenario, there are usually a large number of devices. If actions of all devices are selected in a common DRL, it will inevitably lead to a huge space dimension. For the proposed adaptive blocklength framework, optimizing the blocklength of one device will also affect the optimization of other devices. Therefore, multi-agent DRL can be used to solve the proposed average over-the-air delay minimization problem. We transform the proposed average over-the-air delay minimization problem $\textbf{\textit{P}1}$ into MDP based on the multi-agent DRL. 

In particular, deep Q-network (DQN) is one of the potential value-based DRL methods. DQN refers to the Q-learning algorithm based on deep learning, which mainly uses the value function approximation and neural network (NN) technology. Combining Q-learning with deep learning, DQN can be used to solve the problem of high computational complexity due to the space growth in Q-table. DQN introduces two powerful tools, namely experience replay and Q-target network, which make the NN more efficient and enhance the stability of the algorithm. Specifically, DQN is an offline method, where the corresponding network has been well trained before using it. With trained network, at the beginning of each period, the optimal TTI design and bandwidth allocation in $\textbf{\textit{P}1}$ can be quickly obtained. Thus, DQN can be well applicable to the mURLLC system.

\begin{figure*}[htbp]
	\centering\includegraphics[width=6.3in]{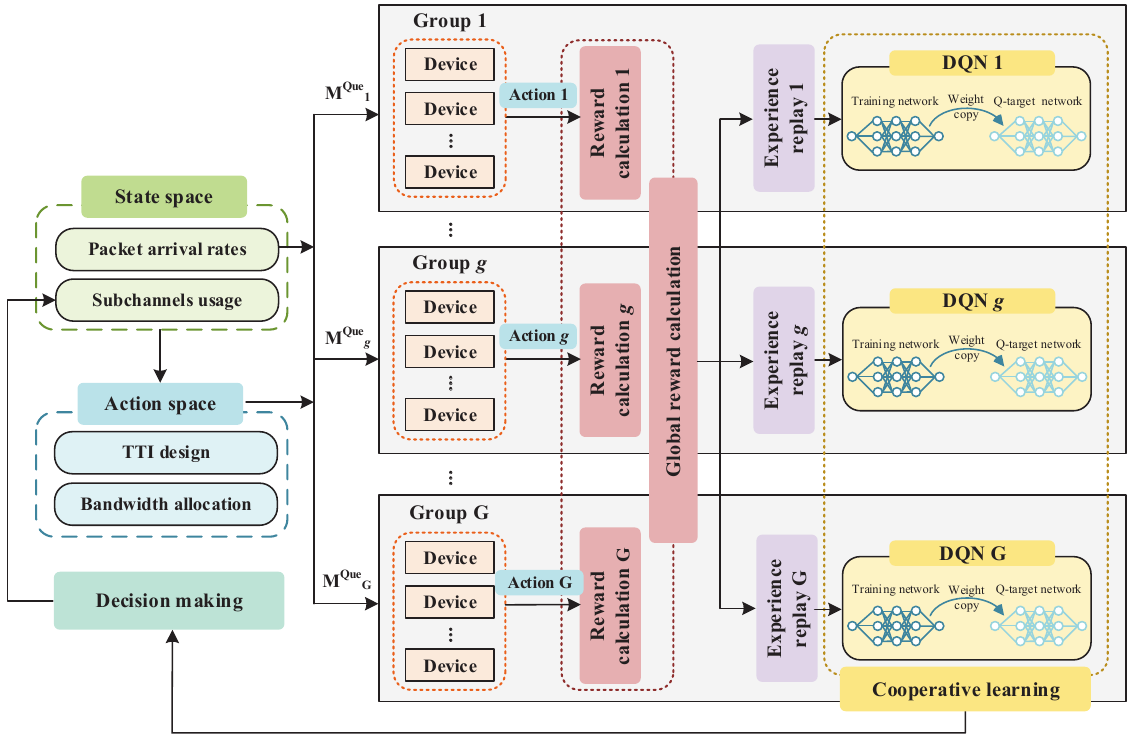}
	\caption{The multi-agent DQN structure for the average over-the-air delay minimization problem.}\label{fig:DQN}
\end{figure*}

Therefore, we use the multi-agent DQN to solve the proposed average over-the-air delay minimization problem. Each agent learns to improve its own policy through interaction with the state in the environment to obtain the optimal policy and maximize the long-term reward of the system, thus reducing the average over-the-air delay. Fig.~\ref{fig:DQN} shows the structure of multi-agent DQN scheme, which contains the state space module, the action selection module, the reward calculation module, the experience replay module, and the DQN module. A detailed explanation of each module for the average over-the-air delay minimization problem is described in the following.

\subsubsection{State Space}
The natural factors in $\textbf{\textit{P}1}$ include the packet arrival rate of each device and the number of subchannels that can be selected. Specifically, the packet arrival rates affect the state transition probability, while the number of remaining subchannels changes when devices take actions. Therefore, we choose these two factors to form the state in the environment, where the state set $\mathcal{S}$ can be expressed as
\begin{align}\label{eq:state}
	\mathcal{S} = \left\{\boldsymbol{\lambda}, M^{\rm ReSubc}\right\},
\end{align}
where $ \boldsymbol{\lambda}=\left\{ \lambda_1, \lambda_2,\cdots, \lambda_K\right\}$ denotes the packet arrival rates of all devices and $M^{\rm ReSubc}$ denotes the remaining subchannels after bandwidth allocation.

\subsubsection{Device Grouping}
When the number of devices is very large, assigning the adaptive subchannel and TTI to each device will inevitably bring a huge amount of computation and consume a lot of computing resources. Thus, we propose a device grouping mechanism to decompose the large action space of the multi-agent DQN. 

The mURLLC devices are equipped with sensors to collect data from the environment, where the packet arrival rates $\boldsymbol{\lambda}$ can be observed~\cite{Sensor}. Then, the maximum queuing length of each device can be derived according to~(\ref{maximum_queuing_length}). In the device grouping mechanism, the active devices with the same maximum queuing length are set into one group. After grouping, the number of groups $G$ and the number of devices in each group $K_g \ (1 \leq g \leq G)$ are recorded. The block resources are optimized for each group, where devices within a group are given the same TTI design and an equal share of the bandwidth resources assigned to them.

\subsubsection{Agent}
Based on the device grouping mechanism, each group is regarded as an agent. Each group $g$ has a dedicated DQN that takes its local current observation of the environment as input and outputs the value function.

\subsubsection{Action Space}
For the average over-the-air delay minimization problem, each agent will decide how the TTIs are designed and which subchannels are assigned. Hence, there are two subsets contained in the action space, namely TTI design and bandwidth allocation. We assume that the period duration $T_{\max}$ is discretized into $L$ parts, and thus the action space set of TTI design, denoted by $\mathcal{A^{\rm TTI}}$, can be expressed as
\begin{align}\label{eq:action1}
	\mathcal{A^{\rm TTI}} = \left\{ \frac{T_{\max}}{L},  \frac{2T_{\max}}{L}, \cdots, T_{\max} \right\}.
\end{align}
Similarly, the action space set of bandwidth allocation, denoted by $\mathcal{A^{\rm BA}}$, can be expressed as
\begin{align}\label{eq:action2}
	\mathcal{A^{\rm BA}} =  \left\{1,2,\cdots,M^{\rm Subc}\right\}.
\end{align}
At the $l$-th time-step, each agent captures local observation $s_{g,l}\in \mathcal{S}$ of the environment. Based on $\mathcal{A^{\rm TTI}}$ and $\mathcal{A^{\rm BA}}$, the $g$-th agent takes the action $a_{g,l}  \in \mathcal{A}$ $ ( \mathcal{A} = \mathcal{A^{\rm TTI}} \cup \mathcal{A^{\rm BA}})$, which contains the the selection of TTI and subchannel. In multi-agent DQN, each agent takes an action, contributing to a global action $\boldsymbol{a_l} = \left\{a_{1,l}, \cdots, a_{g,l}, \cdots, a_{G,l}\right\}$.

\subsubsection{Reward}
The optimization goal of $\textbf{\textit{P}1}$ is to minimize the average over-the-air delay of multiple devices. Thus, we choose a monotonically decreasing function with the average over-the-air delay $D^{\rm Ota}_{\rm Ave}$ as the reward function. Also, the constraints~(\ref{Eq:constranit1})-(\ref{Eq:constranit4}) should be satisfied, otherwise the reward is reduced. To satisfy constraints~(\ref{Eq:constranit1}) and (\ref{Eq:constranit4}), the first cost function is set as the transmission failure which results in the period overrun and the unsatisfied minimum data rate requirements. The transmission failure cost function of the $g$-th agent, denoted by $\zeta^{\rm Trans}_g$, can be expressed as
\begin{align}
	\zeta^{\rm Trans}_g \label{eq:reward-con} =	
	\begin{cases}
		0, \!\!\!&\mbox{if~(\ref{Eq:constranit1}) and (\ref{Eq:constranit4}) hold,} \\
		1, &\mbox{otherwise.}
	\end{cases}
\end{align}
The goal of~(\ref{eq:reward-con}) is to indicate whether the mURLLC service is successful or not. If the transmission requirements of all data packets in the $g$-th group are satisfied within the period, we have $\zeta^{\rm Trans}_g=0$. Otherwise, $\zeta^{\rm Trans}_g=1$. Therefore, the reward function of the $g$-th agent at time-step $l$, denoted by $r_{g,l}$, can be expressed as
\begin{equation}
	r_{g,l} =
	\omega_1  D^{\rm Ota}_{g} + \omega_2 \zeta^{\rm Trans}_g, \label{single_reward}
\end{equation}\label{eq:reward-re}
where $\omega_1$ and $\omega_2$ are both negative constants for balancing the utility ($D^{\rm Ota}_{g}$) and cost ($\zeta^{\rm Trans}_g$). Similarly, to satisfy the constraint~(\ref{Eq:constranit3}), the second cost function is set as the subchannels overuse. The subchannels overuse cost function, denoted by $\zeta^{\rm Subc}$, can be expressed as
\begin{align}
	\zeta^{\rm Subc}\label{eq:reward-subc}=	
	\begin{cases}
		0, \!\!\!&\mbox{if} \ M^{\rm ReSubc} \geq 0, \\
		1, &\mbox{otherwise.}
	\end{cases}
\end{align}
The goal of~(\ref{eq:reward-subc}) is to evaluate the subchannel usage situation. If the sum of subchannels selected by all devices exceeds the total number of subchannels, we have $\zeta^{\rm Subc}=1$. Otherwise, $\zeta^{\rm Subc}=0$. The optimization objective is to minimize the average over-the-air delay, which can be approximately equal to maximize the total rewards. Thus, we use a global reward based on the actions of all agents, where all agents cooperatively work to maximize it. Each group calculates its own reward to get the global reward and performs the corresponding global action. The global reward, denoted by $r_l$, can be expressed as follows: 
\begin{equation}
	r_l  =  \sum_{g=1}^{G} r_{g,l}+ \omega_3 \zeta^{\rm Subc}, \label{global_reward}
\end{equation}\label{eq:reward_total}where $\omega_3$ is a negative constant to satisfy the constraint~(\ref{Eq:constranit3}).

\subsubsection{Experience Replay}
When each agent got the global reward $r_{l}$, the environment changes to the next state $s_{l+1}$ and each agent observes its state $s_{g,l+1}$. In order to better combine Q-learning and DNN, the DQN adopts the experience replay, where the tuple $\left(s_{g,l}, a_{g,l}, r_{l}, s_{g,l+1}\right)$ is stored in the replay buffer. In each episode, a number of data are randomly sampled from the replay for training the Q-network.

\begin{algorithm}[htbp]
	\caption{Cooperative M-DQN scheme for solving $\textbf{\textit{P}1}$.}
	\begin{algorithmic}[1]\label{Table}
		\STATE {Initialization: Initialize the experience replay buffer, the weight of training network $\theta$, the weight of Q-target network $\theta'=\theta$, the weight update period $M^{\theta}$, the number of episodes $M^{\rm Inf}$, and the number of time-steps $L$.}
		\FOR {episode $l = 1, 2,\cdots, M^{\rm Inf}$}
		\STATE Reset the environment.
		\FOR {device $k = 1, 2,\cdots, K$}
		\STATE Observe its state $s_{k}$.
		\ENDFOR
		\STATE Group the devices.
		\FOR {time-step $l$ = $1, 2,\cdots, L$ }
		\FOR {group $g = 1, 2,\cdots, G$}
		\STATE Observes its state $s_{g,l}$.
		\STATE Choose the action $a_{g,l}$ based on the $\epsilon$-greedy action policy.
		\STATE Calculate the instant reward $r_{g,l}$ by using~(\ref{single_reward}).
		\ENDFOR
		\STATE All agents take actions $\boldsymbol{a}_l$ and get the global reward $r_l$ by using~(\ref{global_reward}). The state changes to $s_{l+1}$.
		\FOR {group $g = 1, 2,\cdots, G$}
		\STATE Observes its new state $s_{g,l+1}$.
		\STATE Store the tuple $\left(s_{g,l}, a_{g,l}, r_{l}, s_{g,l+1}\right)$ in the experience replay buffer.
		\STATE Randomly sample a minibatch in the experience replay buffer to train the network by minimizing the loss function~(\ref{eq:loss}).
		\STATE Copy the weight of training network $\theta$ to Q-target network $\theta'$ every $M^{\theta}$ steps.
		\ENDFOR
		\ENDFOR
		\ENDFOR
	\end{algorithmic}
\end{algorithm}

\subsubsection{Q-Function}
To maximize the long-term reward, the DQN chooses the action with the maximum Q-value. The temporal-difference (TD) error, denoted by $\delta_{g,l}$, can be expressed as follows:
\begin{equation}
\begin{split}
	\delta_{g,l} = r_{l}+ \gamma \max Q\left(s_{g,l+1},a_{g,l+1}\right) - Q\left(s_{g,l},a_{g,l}\right),
\end{split}
\end{equation}
where $\gamma \in [0,1]$ denotes the discount factor. A large discount factor corresponds to the large weight of future reward, while a small discount factor represents that the current reward is mainly concerned. 
If the action is independently selected according to the local environmental information, each device cannot obtain the bandwidth allocation actions of other devices when they update the environment at the same time-step. Thus, agents cannot manage to achieve the coordination between multiple optimal actions. To mitigate the impact of exploration from other agents and concurrent actions, the Q-value can be calculated from the action-value function with two learning rates. The action-value function, denoted by $Q\left(s_{g,l},a_{g,l}\right)$, can be expressed as follows:
\begin{align}
	Q\left(s_{g,l},{a_{g,l}}\right)\label{eq:Q_function_2}=	
	\begin{cases}
		Q\left(s_{g,l},{a_{g,l}}\right) + \eta \delta_{g,l}, \!\!\!&\mbox{if} \ \delta_{g,l} \geq 0, \\
		Q\left(s_{g,l},{a_{g,l}}\right) + \beta \delta_{g,l},&\mbox{otherwise},
	\end{cases}
\end{align}
where $\eta \in \left[0,1\right]$ and $\beta\in \left[0,1\right]$ denote two learning rates with $\eta > \beta$. When the Q-value is reduced after the network updating, a smaller learning rate $\beta$ is used. That is, we tolerate the small TD error that resulted in the exploration of other agents, thus allowing other agents to fully explore. With the help of different learning rates, the agents will not completely ignore the penalty, thus avoiding staying in a sub-optimal equilibrium or being uncoordinated on the optimal joint action. 
\subsubsection{Loss Function}
At each training step, each DQN agent updates its weight to minimize the loss function. The loss function, denoted by $L({\theta_{g,l}})$, can be expressed as follows:
\begin{equation}\label{eq:loss}
	\begin{split}
	L({\theta_{g,l}}) = \Big[r_{l} + \gamma \max_{a_{g,l+1}}  Q &  \left(s_{g,l+1},a_{g,l+1}, \theta_{g,l}'\right) \\ &\hspace{0.8cm} - Q\left(s_{g,l},a_{g,l}, \theta_{g,l}\right)\Big]^2,
	\end{split}
\end{equation}
where ${\theta_{g,l}}$ and ${\theta_{g,l}}'$ denote the weights of training network and Q-target network, respectively. After iterations, $\theta_{g,l}$ in the training network is copied to the Q-target network, which makes the Q-target network gradually converge.

\subsubsection{Action Policy}
When the optimal action-value function $Q \left(s_{g,l},{a_{g,l}}\right)$ is achieved, the optimal policy is given by
\begin{equation}\label{Eq:cooperative_optimal_policy}
	\psi_{g,l} = \mathop{\arg \max}\limits_{a_{g,l}} \ Q \left(s_{g,l},{a_{g,l}}\right).
\end{equation}
The behavior selection of the agent is based on $\epsilon$-greedy action policy, where $\epsilon$ represents the exploration probability. In detail, we choose an action of the optimal policy $\psi_{g,l}$ with probability $\epsilon$, while choosing a random action with probability $1-\epsilon$. Finally, using the backpropagation algorithm, the weights of the internal neurons can be updated. The detail of the cooperative M-DQN scheme with the grouping mechanism is given in Algorithm~\ref{Table}. 

\section{Numerical Results}\label{sec:result}
In this section, we evaluate the performance of our proposed adaptive blocklength scheme for mURLLC low-delay communications. We consider a single-cell cellular network with a cell radius of $500$~m, where $K$ devices are randomly distributed with random packet arrival rates $\lambda_k \in ( 0 , 1 ) \ (1 \leq k \leq K)$. We set the inversion power control threshold $P_{0} = -80$~dBm, the noise variance $\sigma^2 = -90$~dBm, the period duration $T_{\max} = 5$~ms, the number of preambles $M^{\rm Pre} = 500$, and the number of subchannels $M^{\rm Subc} = 2000$ with the bandwidth $W=100$~KHz per subchannel, respectively. Specifically, the TTIs of LTE and 5G NR are set as $1$~ms and $0.5$~ms. The processing delay $D^{\rm Proc}$ is set to $10$~$\upmu$s with high-speed routers~\cite{ProcessingDelay}, and the propagation delay $D^{\rm Prop} $ is determined by the distance from each device to the BS. The learning rates $\eta$ and $\beta$ are set to $0.01$ and $0.001$. The discount factor $\gamma$ is set to $0.9$ and the random action probability $\epsilon$ is set from $0.99$ to $0.05$. In addition, the parameters $\omega_1$ and $\omega_2$ used to calculate the reward are set to $-1000$ and $-1$, and $\omega_3$ is determined by the device grouping situation.

\begin{figure}[htbp] 
	\centering  
	\vspace{-0.35cm} 
	\subfigtopskip=2pt 
	\subfigbottomskip=2pt 
	\subfigcapskip=-5pt 
	\subfigure[The successful access probability versus the number of bits per packet.]{
		\label{Fig:bitperPacket_psuc}
		\includegraphics[width=0.9\linewidth]{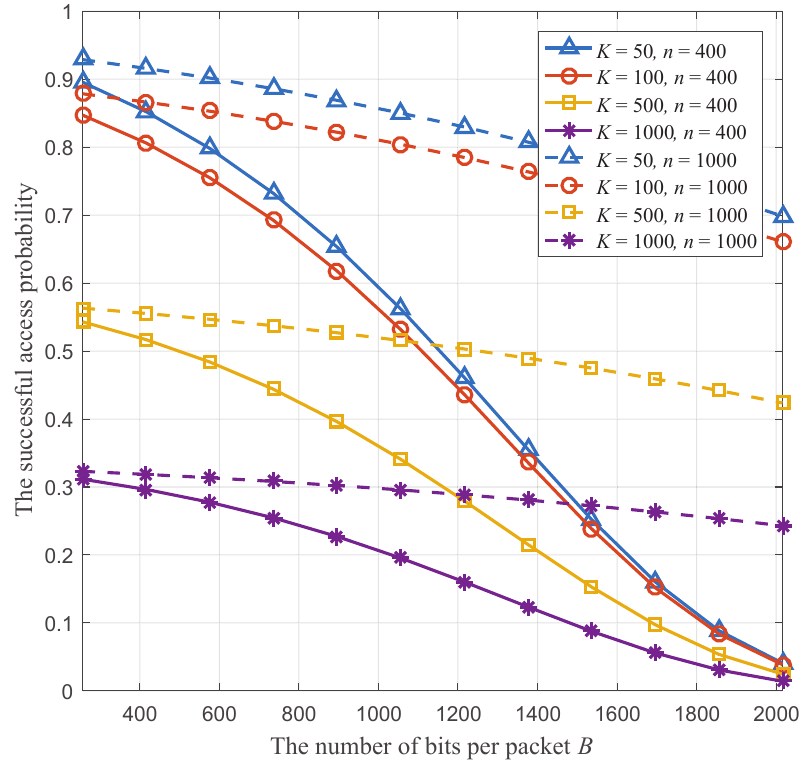}}
	\quad 
	\subfigure[The number of retransmissions versus the number of bits per packet.]{
		\label{Fig:bitperPacket_retransmission}
		\includegraphics[width=0.9\linewidth]{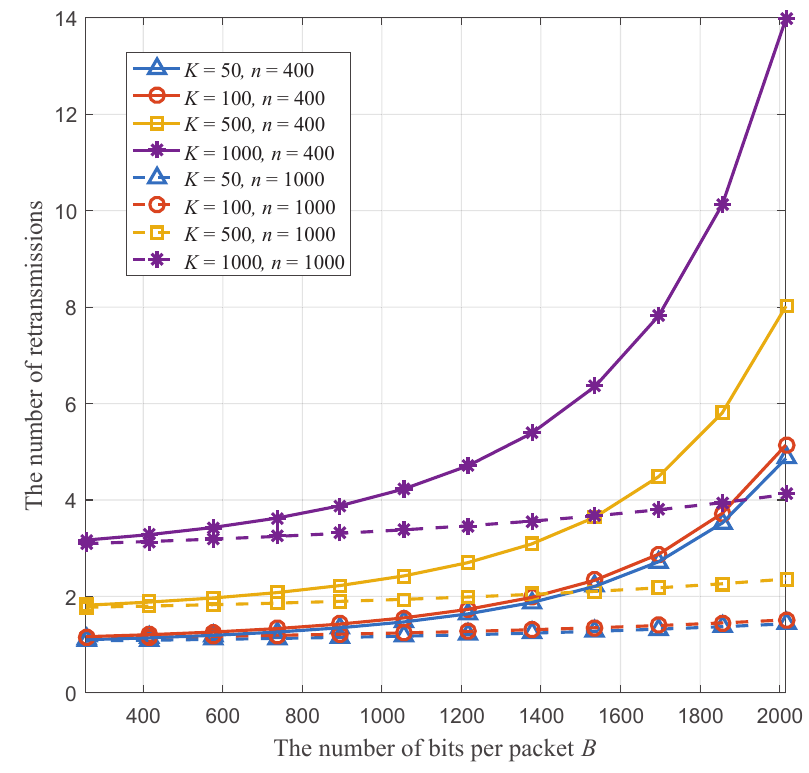}}
	\caption{Performance evaluation in terms of the number of bits per packet.}
	\label{Fig:bitperPacket}
\end{figure}
Figure~\ref{Fig:bitperPacket} depicts the two performance evaluation indicators, i.e., the successful access probability and the number of retransmissions, versus the number of bits per packet under different blocklengths and numbers of devices. Overall, the successful access probability decreases and the number of retransmissions increases as the number of bits per packet increases. When the number of bits in each packet is large, the successful access probability approaches zero with a short blocklength. This means that variable blocklengths are required to adapt to packets with different numbers of bits, thus avoiding a large number of retransmissions.

\begin{figure}[htbp] 
	\centering 
	\vspace{-0.35cm} 
	\subfigtopskip=2pt 
	\subfigbottomskip=2pt 
	\subfigcapskip=-5pt 
	\subfigure[The successful access probability versus the number of devices.]{
		\label{Fig:KUser_psuc}
		\includegraphics[width=0.9\linewidth]{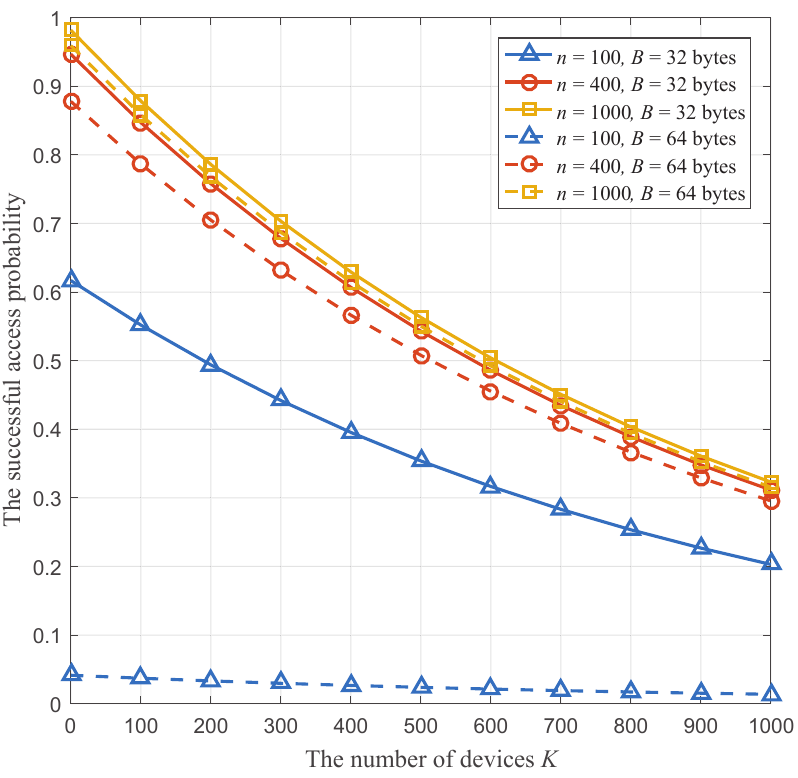}}
	\quad
	\subfigure[The number of retransmissions versus the number of devices.]{
		\label{Fig:KUser_retransmission}
		\includegraphics[width=0.9\linewidth]{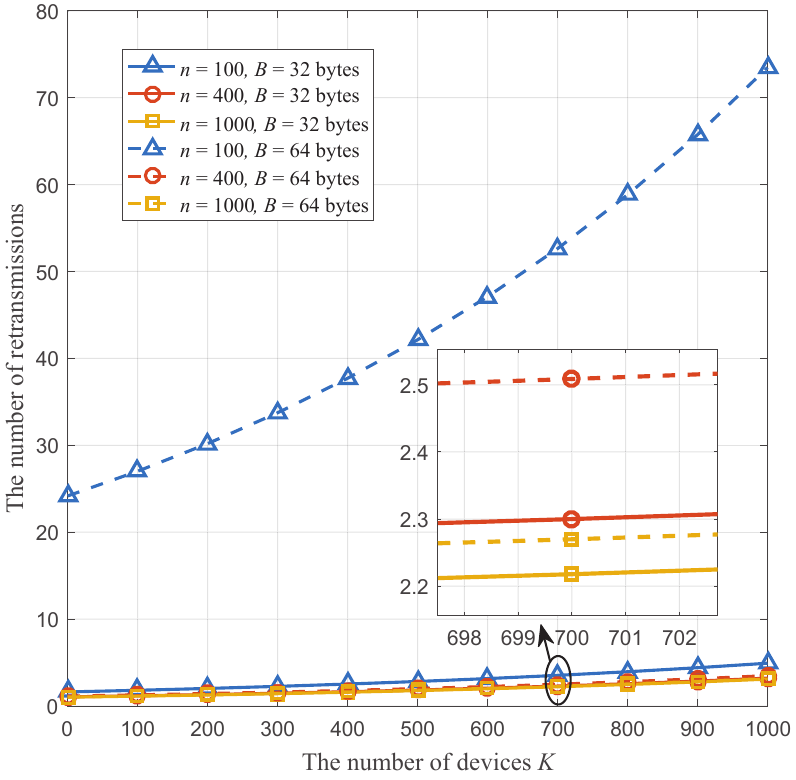}}
	\caption{Performance evaluation in terms of the number of devices.}
	\label{Fig:KUser}
\end{figure}
Figure~\ref{Fig:KUser} shows the above two performance evaluation indicators versus the number of devices under different blocklengths and numbers of bits. Similar to the trend shown in Fig.~\ref{Fig:bitperPacket}, the successful access probability gradually decreases and the number of retransmissions gradually increases as the number of devices increases. 
Given the same number of bits per packet, the increase of the successful access probability for the blocklength from $400$ to $1000$ is smaller than the increase of the successful access probability for the blocklength from $100$ to $400$. In addition, when the blocklength is relatively large, the number of bits per packet has little impact on the successful access probability and the number of retransmissions. This indicates that although the large blocklength leads to a large transmission delay, it can guarantee the reliability of transmission with heavy data load.

\begin{figure}[htbp]
	\centering 
	\vspace{-0.35cm} 
	\subfigtopskip=2pt 
	\subfigbottomskip=2pt 
	\subfigcapskip=-5pt
	\subfigure[The successful access probability versus the blocklength.]{
		\label{Fig:blocklength_psuc}
		\includegraphics[width=0.92\linewidth]{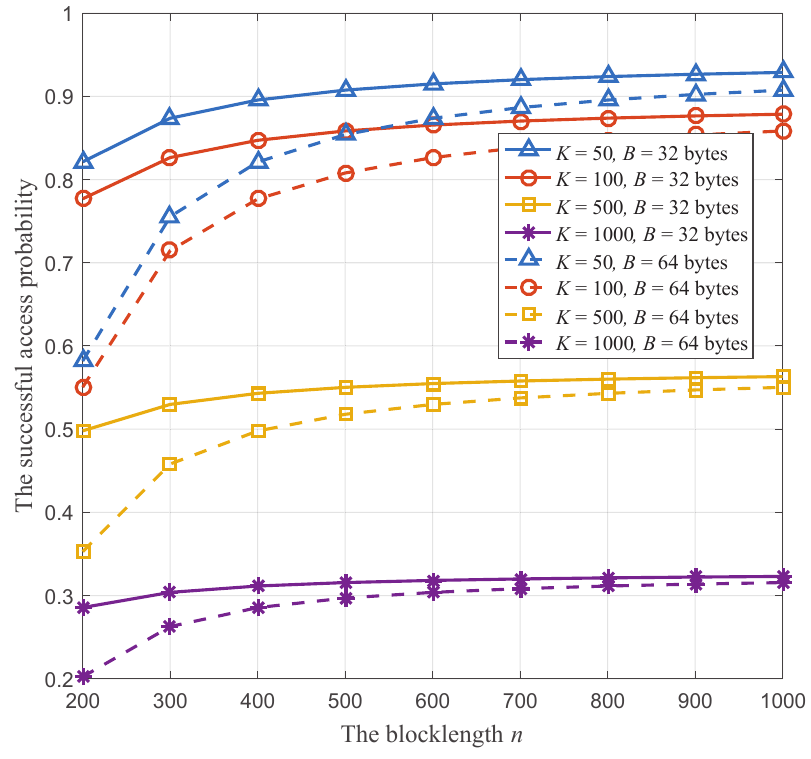}}
	\quad
	\subfigure[The number of retransmissions versus the blocklength.]{
		\label{Fig:blocklength_retransmission}
		\includegraphics[width=0.92\linewidth]{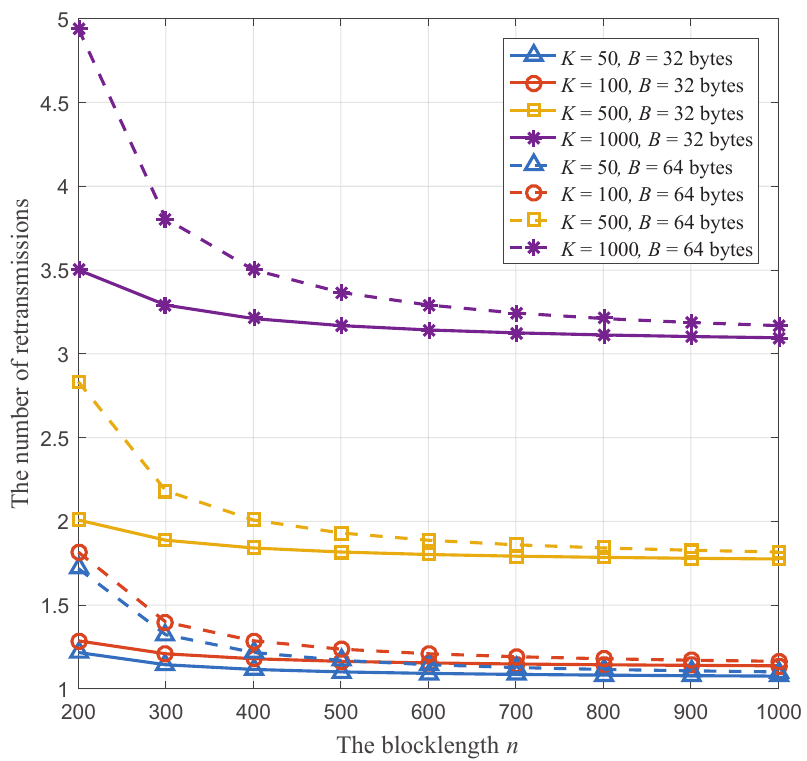}}
	\caption{Performance evaluation in terms of the blocklength.}
	\label{Fig:blocklength}
\end{figure}
Figure~\ref{Fig:blocklength} plots the two performance evaluation indicators versus the blocklength under different numbers of devices and bits. The successful access probability monotonically increases and the number of retransmissions monotonically decreases as the blocklength increases. It is worth noting that when the blocklength exceeds $600$, the number of retransmissions almost keeps unchanged with the increase of the blocklength. In this case, only the queuing delay and the transmission delay affect the tradeoff relationship. 

\begin{figure}[htbp]
	\centering\includegraphics[width=3.25in]{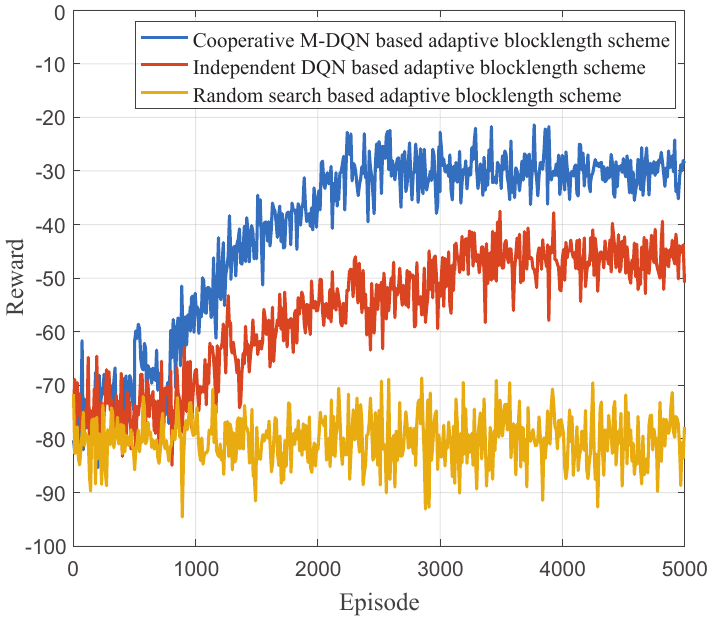}
	\caption{Reward comparison of the adaptive blocklength schemes.}\label{fig:Convergence}
\end{figure}
Figure~\ref{fig:Convergence} compares the rewards of the proposed cooperative M-DQN scheme, the independent DQN scheme, and the random search scheme as the training episode increases to investigate the convergence behavior. It can be seen from Fig.~\ref{fig:Convergence} that the proposed cooperative M-DQN scheme achieves a higher reward than that of the independent DQN scheme and the random search scheme, which means that the proposed cooperative M-DQN scheme converges to the optimal policy. In addition, adopting the grouping mechanism, the proposed cooperative M-DQN scheme has a faster convergence speed compared with that of the independent DQN scheme and the random search scheme. Without any cooperation between devices, the independent DQN scheme lacks consideration of global performance, resulting in a lower reward. Although the random search scheme has the simplest structure, the worst reward is achieved and kept unchanged as the episode increases. The proposed cooperative M-DQN scheme with a grouping mechanism can enhance the convergence speed and learning efficiency, which is very useful to solve the average over-the-air delay minimization problem.

\begin{figure}[htbp]
	\centering\includegraphics[width=3.2in]{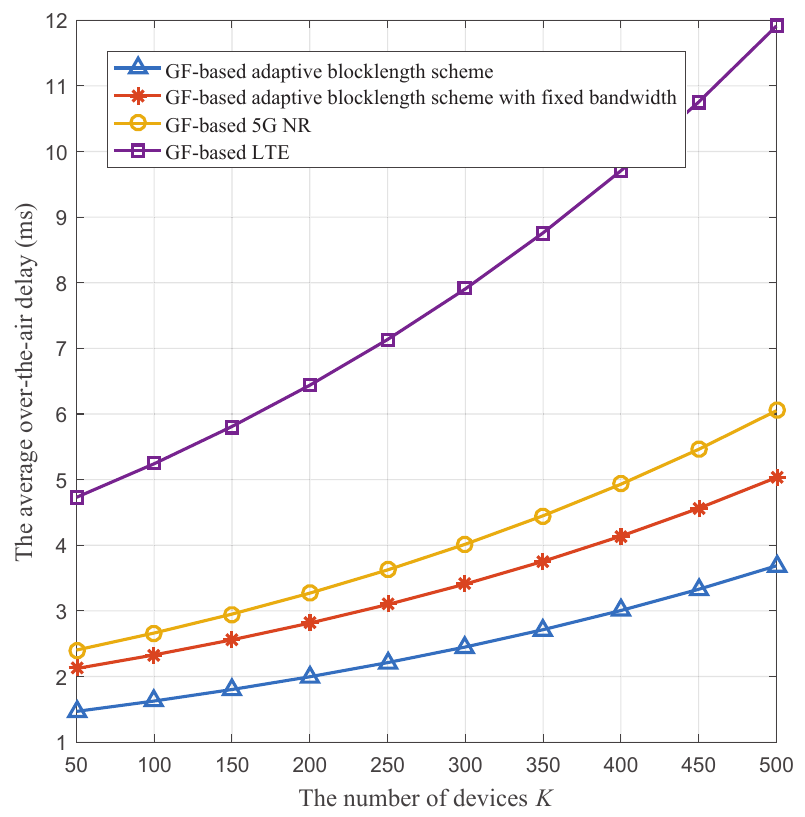}
	\caption{The over-the-air delay of the proposed adaptive blocklength scheme versus the number of devices compared with that of LTE and 5G NR.}\label{fig:AD_K}
\end{figure}

Figure~\ref{fig:AD_K} depicts the average over-the-air delay of the proposed adaptive blocklength scheme versus the number of devices compared with that of LTE and 5G NR. Due to the mismatch between the numbers of devices and preambles as well as the limited bandwidth resource, the average over-the-air delays of the proposed adaptive blocklength scheme, LTE, and 5G NR accordingly increase as the number of devices increases. Given the same number of devices, our proposed adaptive blocklength scheme can achieve the minimum average over-the-air delay to transmit all of the arrival packets compared with that of LTE and 5G NR. This proves that our proposed adaptive blocklength scheme can effectively reduce the over-the-air delay. In addition, even though the average over-the-air delays of the three schemes increase as the number of devices increases, the growth rate of the proposed adaptive blocklength scheme is the lowest, which means the blocklength can be adaptively changed in time to adapt to the real-time load when the number of devices changes. Also, it can be observed from Fig.~\ref{fig:AD_K} that if only changing the TTI under the fixed bandwidth allocation, the over-the-air delay is higher than that of the proposed scheme with adaptive bandwidth allocation. With the joint TTI and bandwidth adaption, the proposed adaptive blocklength scheme can achieve the minimum average over-the-air delay. 

\begin{figure}[htbp]
	\centering\includegraphics[width=3.2in]{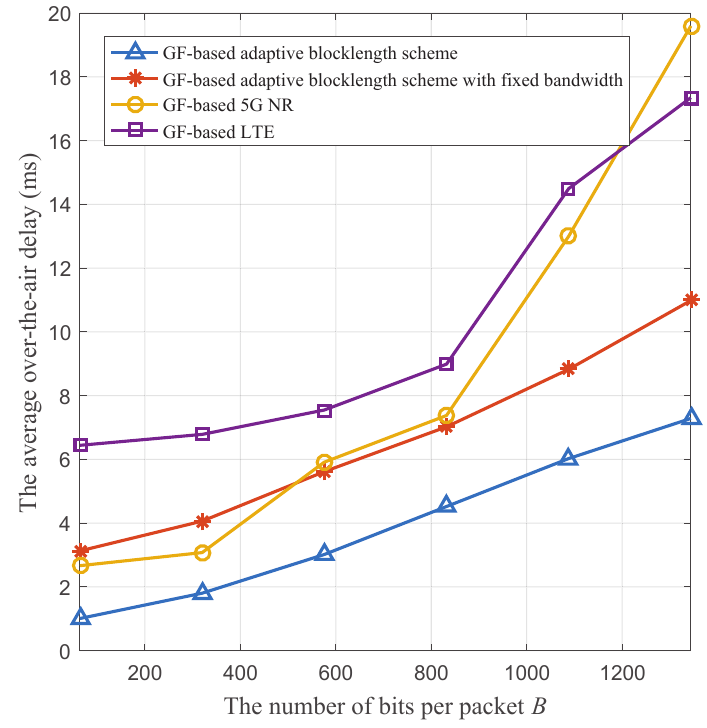}
	\caption{The over-the-air delay of the proposed adaptive blocklength scheme versus the number of bits per packet compared with that of LTE and 5G NR.}\label{fig:AD_B}
\end{figure}

Figure~\ref{fig:AD_B} plots the average over-the-air delay of the proposed adaptive blocklength scheme versus the number of bits per short packet compared with that of LTE and 5G NR. The over-the-air delays of the three schemes increase as the number of bits per packet increases. Given the same number of bits per packet, our proposed adaptive blocklength scheme can achieve the minimum average over-the-air delay compared with that of LTE and 5G NR. It is worth mentioning that when the number of bits per packet is relatively large, the average over-the-air delay of 5G NR is larger than that of LTE. This is because when the number of bits is large, a short blocklength structure in 5G NR cannot completely transmit all the data of one packet, leading to an increase in the number of required blocks. Although the transmission delay of short blocklength in 5G NR is smaller than that of LTE with long blocklength, the queuing delay increases as the number of blocks increases. The proposed adaptive blocklength scheme can adaptively change the blocklength according to the specific traffic load, which avoids this situation and achieve the optimal tradeoff to obtain the minimum over-the-air delay. Specifically, when the numbers of bits corresponding to packets are different, the adaptive blocklength scheme can flexibly change the TTI, avoiding the problem of large transmission delay when transmitting short packets in LTE and the problem of large queuing delay when transmitting long packets in 5G NR with traditional fixed TTI. In addition, we can observe that if only the TTI is adaptively changed, the corresponding delay performance is possibly worse than that of 5G NR when the number of bits per packet is relatively small. This indicates the need for adaptive bandwidth allocation, which is integrated into our proposed adaptive blocklength scheme.

\section{Conclusion}\label{sec:con}
In this paper, we proposed the adaptive finite blocklength scheme and solved the problem of how to reduce the over-the-air delay for mURLLC in 6G wireless networks. Taking the short packet into account, we revealed the tradeoff relationship among the queuing delay, the transmission delay, and the number of retransmissions along with the change of finite blocklength, based on which we formulated the adaptive blocklength framework. Then, analytical expressions were derived for the successful access probability and the steady-state probability distribution, where the queuing state update model of the GF-based adaptive blocklength framework was correspondingly established. On this basis, we formulated the average over-the-air delay minimization problem to jointly design the TTI and allocate the bandwidth. Finally, we developed the cooperative multi-agent deep Q-network scheme with a grouping mechanism to efficiently deal with this non-convex problem, solving which we can obtain the adaptive blocklength scheme. Compared with the fixed blocklength structure in LTE and 5G NR systems, our proposed adaptive blocklength scheme can significantly reduce the over-the-air delay. 

\bibliographystyle{IEEEtran}
\bibliography{IEEEabrv,References}

\vfill

\end{document}